\documentclass[12pt]{article}
\usepackage{a4wide,epsfig}

\newcommand{\logminus}{\ln\left(2-z^{\beta_0}\right)}
\newcommand{\betnull}{\beta_0}
\newcommand{\mrM}{\mu_r}

\newcommand{\bfm}[1]{\mbox{\boldmath$#1$}}

\def\als{\alpha_{\rm s}}

\newcommand{\nn}{\nonumber}
\newcommand{\be}{\begin{equation}}
\newcommand{\ee}{\end{equation}}
\newcommand{\bea}{\begin{eqnarray}}
\newcommand{\eea}{\end{eqnarray}}

\begin{document}

\title{\vskip-3cm{\baselineskip14pt
\centerline{\normalsize DESY 04-042\hfill}
\centerline{\normalsize TTP04-06 \hfill}
\centerline{\normalsize UB-ECM-PF-04-05 \hfill}
}
\vskip1.5cm
$M(B^*_c)-M(B_c)$  Splitting from Nonrelativistic Renormalization Group}
\author{A.A. Penin$^{a,b}$, A. Pineda$^c$,
V.A. Smirnov$^{d,e}$,  M. Steinhauser$^e$\\
{\small\it $^a$ Institut f{\"u}r Theoretische Teilchenphysik,
Universit{\"a}t Karlsruhe, 76128 Karlsruhe, Germany}\\
{\small\it $^b$ Institute for Nuclear Research, Russian Academy of Sciences,
117312 Moscow, Russia}\\
{\small\it $^c$ Dept. d'Estructura i Constituents de la Mat\`eria,
  U. Barcelona,
E-08028 Barcelona, 
Spain}\\
{\small\it $^d$ Institute for Nuclear Physics, Moscow State
  University, 119992 Moscow, Russia}\\
{\small\it $^e$ II. Institut f\"ur Theoretische Physik, Universit\"at Hamburg,
22761 Hamburg, Germany}}

\date{}

\maketitle

\thispagestyle{empty}

\begin{abstract}
We compute the hyperfine splitting in a heavy quarkonium composed of different
flavors in next-to-leading logarithmic approximation using the nonrelativistic
renormalization group.  We predict the mass difference of the vector and
pseudoscalar charm-bottom mesons to be $M(B^*_c)-M(B_c)=46 \pm 15\,{(\rm
th)}\,{}^{+13}_{-11}\,(\delta\alpha_s)$~MeV.
\medskip

\end{abstract}



\section{Introduction}

The recently discovered charm-bottom heavy quarkonium completes the well
investigated charmonium and bottomonium families and offers a new perspective
in the study of the nonrelativistic dynamics of the strong interactions. The
first experimental observation of about twenty events interpreted as the
decays of the $B_c$ meson by CDF collaboration \cite{Abe} does not match the
precision of the spin one charmonium and bottomonium measurements.  The
statistics, however, is expected to increase significantly in future
experiments at Tevatron and LHC greatly improving the accuracy of the
data.  Note that only the pseudoscalar (spin singlet) state has been observed
so far while the vector (spin triplet) meson $B_c^*$ is still to be
discovered. This distinguishes $c\bar b$ quarkonium from the $b\bar b$ system,
where it is the pseudoscalar $\eta_b$ meson, which asks for experimental
detection.

From the theoretical point of view, the charm-bottom mesons are ``in between''
the approximately Coulomb $b\bar b$ mesons and the $c\bar c$
mesons. Therefore, a simultaneous analysis of all three quarkonia could shed
new light on the balance between the perturbative and nonperturbative effects
and further check whether a perturbative analysis provides a reliable starting
point for them. Moreover, since the nonperturbative effects in the $c\bar b$
system are suppressed with respect to the $c\bar c$ meson, the former could be
a cleaner place to determine the charm quark mass (provided the experimental
accuracy is good enough). Another point to be stressed is that, though the
leading order dynamics of the $c\bar b$ state is quite similar to the $b\bar
b$ and $c\bar c$ one (up to the value of the reduced mass) the higher order
relativistic and perturbative corrections are different. Thus the comparison
of $c\bar b$ and $b\bar b$ ($c\bar c$) properties could help to establish fine
details of the nonrelativistic dynamics.

The spectrum of the charm-bottom quarkonium has been subject of numerous
investigations based on potential models
\cite{Eichten:1994gt,Gershtein:1994jw}, lattice simulations \cite{Sha}, and
pNRQCD \cite{BraVai}. This last analysis computed the ground state energy
within a pure perturbative approach. We consider that this analysis further
indicates that a perturbative approach can be a good starting point for
studing the $B_c$ system.

In the present paper we focus on the hyperfine splitting (HFS) $E_{\rm hfs}$
of the $B_c$, {\it i.e.} the mass difference between the singlet and triplet
spin states $M(B^*_c)-M(B_c)$. The QCD study of the heavy quarkonium HFS has a
long history \cite{BucNg,recent}.  For the same-flavor quarkonium the
next-to-leading order (NLO) ${\cal O}(\alpha_s)$ correction to the ground
state HFS can be found in \cite{PenSte} in a closed analytical form.  The
leading order HFS is proportional to the fourth power of the strong coupling
constant $\alpha_s(\nu)$ and thus the low order calculations suffer from
strong spurious dependence on the renormalization scale $\nu$, which
essentially limits the numerical accuracy of the approximation.  Hence, the
proper fixing of the normalization scale becomes mandatory for the HFS
phenomenology.  The dynamics of the nonrelativistic bound state, however, is
characterized by three well separated scales: the hard scale of the heavy
quark mass $m$, the soft scale of the bound state momentum $mv$, and the
ultrasoft scale of the bound state energy $mv^2$, where $v\propto\alpha_s$ is
the velocity of the heavy quark inside the approximately Coulomb bound state.
To make the procedure of scale fixing self-consistent one has to resum to all
orders the large logarithms of the scale ratios.  For the same-flavor case
this problem has been solved in Ref.~\cite{KPPSS} within the nonrelativistic
renormalization group (NRG) approach and the next-to-leading logarithmic (NLL)
result for HFS has been derived.  The renormalization group improved result
shows better stability with respect to the scale variation.  Moreover, the use
of the NRG significantly improves the agreement with the experimental value of
HFS in charmonium in comparison to the NLO computation.  Below we generalize
the analysis to the different-flavor quarkonium case and apply the result to
predict the splitting $M(B^*_c)-M(B_c)$.


\section{Renormalization group running of the spin dependent potential}

To derive the NRG equations necessary for the NLL analysis of the HFS, we rely
on the method based on the formulation of the nonrelativistic effective theory
\cite{CasLep} known as potential NRQCD (pNRQCD) \cite{PinSot1}. The method was
developed in Ref.~\cite{Pin} where, in particular, the leading logarithmic
(LL) result for HFS has been obtained (see also Ref.~\cite{HMS}).  In pNRQCD
the HFS is generated by the spin-flip potential in the effective Hamiltonian,
which in momentum space has the form
\be 
\delta {\cal  H}_{\rm spin}  = D_{S^2,s}^{(2)}{4C_F\pi\over 3m_1m_2}\bfm{S}^2,
\qquad {\bfm S}={{\bfm \sigma}_1+{\bfm \sigma}_2\over 2}\,,
\label{pot}
\ee
where ${\bfm \sigma}_{1}$ and ${\bfm \sigma}_{2}$ are the spin operators of
the quark and antiquark with masses $m_1$ and $m_2$,
$C_F=(N_c^2-1)/(2N_c)$, and $D_{S^2,s}^{(2)}$ is the Wilson coefficient, which
incorporates the effects of the modes that have been integrated out.  In
effective theory calculations such couplings become singular as a result of
the scale separation. The renormalization of these singularities allows one
to derive the equations of the NRG,
which describe the running of the effective-theory couplings, {\it i.e.} their
dependence on the effective-theory cutoffs.  The solution of these equations
sums up the logarithms of the scale ratios. 

In general, one should consider
the soft, potential and ultrasoft running of $D_{S^2,s}^{(2)}$ 
corresponding to the ultraviolet
divergences of the soft, potential, and ultrasoft regions \cite{BenSmi}. We
denote the corresponding cutoffs as $\nu_s$, $\nu_p$ and $\nu_{us}$,
respectively. $\nu_{us}$ and $\nu_{p}$ are correlated as was first realized in Ref. \cite{LMR}. 
A natural relation between them is 
$\nu_{us}=\nu_p^2/(2m_r)$, where $m_r=m_1m_2/(m_1+m_2)$ is the reduced mass.
The dependence on $\nu_s$ first emerges 
in  the LL approximation after integrating out the hard modes. 
It disappears after subsequent integrating out the soft modes
giving rise to a dependence on  $k$,  the
three-dimensional momentum transfer  between the quark and antiquark. 
Thus the soft running  effectively stops at $\nu_s=k$.
The dependence on $\nu_p$ emerges for the first time in the NLL approximation
and cancels out  in the time-independent Schr\"odinger perturbation
theory for heavy quarkonium observables. Thus, in pNRQCD one considers 
$D_{S^2,s}^{(2)}$ as a function of $k$ and $\nu_p$. 
For the calculation of the spectrum it is convenient to expand this 
$k$-dependent potential around $k=\nu_s$ 
\be
D_{S^2,s}^{(2)}(k,\nu_p)=D_{S^2,s}^{(2)}(\nu_s,\nu_p)+
\ln\left({k\over\nu_s}\right)
\nu_s{d \over d\nu_s} D_{S^2,s}^{(2)}(\nu_s,\nu_p)+\ldots
\,.
\label{momdep}
\ee
The characteristic momentum for the Coulomb system
is $ \alpha_sm_r$ and for $\nu_s\sim \alpha_sm_r$ 
the average of  $\ln\left({k/\nu_s}\right)$
over  bound state wave function does not produce a large logarithm
while the derivative in $\ln \nu_s$ results in extra factor of $\alpha_s$. 
Thus, for the calculation  of the HFS in NLL approximation 
one can take the first term  on the right hand side of
Eq.~(\ref{momdep}) in the NLL approximation, the second term in the LL
approximation and neglect the higher derivative terms.

Once expanded, the potential is a function of $\nu_s$ and $\nu_p$ (we should
not forget that there is also a dependence on $m_i$, the masses of the heavy
quarks, and $\nu_h$, the matching scale of the order of the heavy quark
masses).  Let us start with the discussion of the soft running.  To the NLL
approximation it is determined by the following NRG equation
\begin{eqnarray}
  \nu_s {d\over d\nu_s}D_{S^2,s}^{(2)}
  &=&
  \alpha_s
  c_F(m_1)c_F(m_2)\gamma_s
  \label{Dsoftrun}
  \,,
\end{eqnarray}
where $c_F$ is the  effective Fermi coupling,  
\begin{eqnarray}
  \gamma_s&=&\gamma^{(1)}_s{\alpha_s \over
    \pi}+
  \gamma^{(2)}_s{\alpha_s^2 \over \pi^2}+\cdots
\end{eqnarray}
is the soft anomalous dimension and $\alpha_s = \alpha_s(\nu_s)$ is
renormalized in the $\overline{\rm MS}$ scheme. The running of the
coefficient $c_F$ is known  in  NLL
approximation  \cite{ABN}. It reads 
\bea
c_F(m_i)&=& 
z^{-{\gamma_0\over 2}}
\left[ 1 + \frac{\alpha_s(\nu_h)}{4\pi}
  \left(c_1+\frac{\gamma_0}{2}\ln\frac{\nu_h^2}{m_i^2}\right) 
  + \frac{\alpha_s(\nu_h) - \alpha_s(\nu_s)}{4\pi}\left(
    \frac{\gamma_1}{2\beta_0} - \frac{\gamma_0\beta_1}{2\beta_0^2}
  \right) + \dots \right] 
\,,
\nonumber\\
\eea
where  $z=\left(\alpha_s(\nu_s)/\alpha_s(\nu_h)\right)^{1/\beta_0}$, 
$\nu_h\sim m_i$ is the hard matching scale,  $c_1 = 2(C_A+C_F)$ 
and the one- and two-loop   anomalous dimensions read~\cite{ABN}
\be
   \gamma_0 = 2 C_A \,, \qquad
   \gamma_1 = \frac{68}{9}\,C_A^2 - \frac{52}{9}\,C_A T_F\,n_l
   \,.
\ee
Here  $C_A=N_c$, $T_F=1/2$,   
$n_l$ is the number of massless quark flavors, 
and $\beta_i$ is the $(i+1)$-loop
coefficient  of the QCD $\beta$ function
\be
  \beta_0 = \frac{11}{3}C_A-\frac{4}{3} T_Fn_l\,, \qquad
  \beta_1 = \frac{34}{3}C_A^2-\frac{20}{3} C_A T_Fn_l - 4C_FT_Fn_l\,.
\ee
The value of one-loop anomalous dimension
\begin{eqnarray}
  \gamma^{(1)}_s &=& -\frac{\beta_0}{2} + \frac{7}{4}C_A
\end{eqnarray}
can be extracted  from the result of Ref.~\cite{Pin}. 
The result for the two-loop coefficient
\begin{eqnarray}
  \gamma^{(2)}_s&=&
  \frac{1}{216}
  \left[
    {{C_A}}^2\,\left( 5 - 36\,{\pi }^2 \right)  + 
    88\,{C_A}\,{n_l}\,{T_F} + 
    4\,{n_l}\,{T_F}\,\left( 27\,{C_F} - 
      40\,{n_l}\,{T_F} \right)
  \right]
  \,,
\label{gamma2}
\end{eqnarray}
is new. It was obtained by an explicit calculation of the subleading
singularities of the two-loop soft diagrams using the approach of
\cite{PinSot2,CMY,KPSS}.  In this approach, dimensional regularization with
$D=4-2\varepsilon$ is used to handle the divergences, and the formal
expressions derived from the Feynman rules of the effective theory are
understood in the sense of the threshold expansion \cite{BenSmi}.  Thus the
practical calculation reduces to the evaluation of the coefficients of the
quadratic and linear soft poles in $\varepsilon$.  Our approach possesses two
crucial virtues: the absence of additional regulator scales and the automatic
matching of the contributions from different scales.  For the reduction of the
two-loop Feynman integrals to the master ones the method of Ref.~\cite{Bai}
was used.

The solution of  Eq.~(\ref{Dsoftrun}) can be written as a sum of the LL
and NLL contributions.
The LL result  is already known and reads \cite{Pin}
(see also \cite{HMS})
\be
\left(D_{S^2,s}^{(2)}\right)^{LL}=\alpha_s(\nu_h)
\left[
1
+{2\beta_0-7C_A \over 2\beta_0-4C_A}\left(
 z^{-2C_A+\beta_0} -1 \right) \right]\,.
\label{DLL}
\ee
For the NLL term we obtain
\be
\left(\delta D_{S^2,s}^{(2)}\right)^{NLL}_{s}
=B_1\alpha_s^2(\nu_h)(z^{-\gamma_0+\beta_0}-1)+
B_2\alpha_s^2(\nu_h)(z^{-\gamma_0+2\beta_0}-1)
\,,
\label{DNLLs}
\ee
where
\bea
B_1&=& \frac{
     {\beta_1}{\gamma_0} - 
     2 \beta_0^2
          \left[c_1
            +{\gamma_0 \over 2}\ln\left(\frac{\nu_h^2}{m_1m_2}\right)\right] 
          - \beta_0\gamma_1}{2
    {{\beta_0}}^2
    \left( {\beta_0} - {\gamma_0} \right) 
    \pi } {\gamma^{(1)}_s}
\,,
\\
B_2&=& \frac{
     - {\beta_1}\,{\gamma_0}\,
         {\gamma^{(1)}_s}  + 
      {\beta_0}\,{\gamma_1}\,
       {\gamma^{(1)}_s} + 
      {{\beta_0}}\,
       \left( {\beta_1}\,{\gamma^{(1)}_s} - 
         4\,\beta_0\,{\gamma^{(2)}_s} \right)  }{2\,
    {{\beta_0}}^2\,
    \left( 2\,{\beta_0} - {\gamma_0} \right)
      \,\pi }
\,.
\eea
The potential running starts to contribute in NLL order. 
To compute it we inspect all operators that lead to spin-dependent
ultraviolet divergences in the time-independent perturbation theory
contribution with  one and two potential loops \cite{Pin,KniPen2,Hil}.
They are
\begin{itemize}
\item[(i)]  the ${\cal O}(v^2,\alpha_sv)$ operators 
  \cite{BucNg}, 
\item[(ii)] the tree ${\cal
    O}(v^4)$ operators, some of which can be checked against the QED analysis
  \cite{CMY,Pac}, 
\item[(iii)] the one-loop ${\cal O}(\alpha_sv^3)$ operators for which
  only the Abelian parts are known \cite{CMY}, while the non-Abelian parts
  are new.
\end{itemize}
In the NLL approximation, we need the LL soft and ultrasoft running of the
${\cal O}(v^2)$ and ${\cal O}(v^4)$ operators, which enter the two-loop
time-independent perturbation theory diagrams, and the NLL soft and ultrasoft
running of the ${\cal O}(\alpha_sv)$ and ${\cal O}(\alpha_sv^3)$ operators,
which contribute at one loop.  The running of the ${\cal O}(v^2,\alpha_sv)$
operators is already known within pNRQCD \cite{Pin}. The running of the other
operators is new. For some of them, it can be obtained using the
reparameterization invariance \cite{Manohar}. 
We refrain from writing the corresponding system of NRG equations, which 
is rather lengthy, and only present its solution, which can be cast in the form
\bea
      \left(\delta D_{S^2,s}^{(2)}\right)^{NLL}_{p}&=&
      \pi \alpha_s^2(\nu_h) \sum_{i=1}^{18} A_i f_i
      \,,
      \label{DNLLp}
\eea
where the coefficients $A_i$ and $f_i$ are given in the Appendix.
To get this result we rescale the ultrasoft cutoff to
$\nu_{us}=\nu_p^2/\nu_h$.  The  difference to  the 
previous definition is beyond the NLL accuracy.

The LL result~(\ref{DLL}) obeys the tree level matching condition 
\be
\left.\left(D_{S^2,s}^{(2)}\right)^{LL}\right|_{\nu=\nu_h}=
\alpha_s(\nu_h) \,,
\ee
while Eqs.~(\ref{DNLLs}) and~(\ref{DNLLp})  vanish at $\nu=\nu_h$ by
construction. 
We then use the known one-loop result of the potential \cite{BucNg} to obtain the NLO
matching condition at the scale $k=\nu_s=\nu_p=\nu_h$. It reads
\bea
    \left(D_{S^2,s}^{(2)}\right)_{\rm 1-loop}&=&
      \left[
        -{5 \over 9}T_Fn_l
        -\frac{5}{36}C_A +C_F+{7 \over 8}C_A\ln\left(\frac{\nu_h^2}{m_1m_2}\right)
      \right.\nn\\&&
      \left.
        -\frac{3}{4}\left(
          C_F\frac{m_1-m_2}{m_1+m_2}
          +\frac{1}{2}\left(C_A-2C_F\right)\frac{m_1+m_2}{m_1-m_2}
        \right)\ln\left(\frac{m_2}{m_1}\right)
      \right]
      {\alpha_s^2(\nu_h) \over \pi}
    \,.
    \nn\\
    \label{DNLL1l}
\eea
Note that in the limit $m_1=m_2\equiv m_q$ this equation does not reproduce 
the same-flavor equal-mass expression
\bea
      \left(D_{S^2,s}^{(2)}\right)^{q\bar q}_{\rm 1-loop}&=&
      \left[
        -{5 \over 9}T_Fn_l+{3 \over 2}(1-\ln 2)T_F+{11C_A-9C_F \over 18}
      \right.\left.
        +{7 \over 4}C_A\ln\frac{\nu_h}{m_q}
      \right]
      {\alpha_s^2(\nu_h) \over \pi}
    \,,\nonumber
\\
\eea
because of the two-gluon annihilation contribution present in the latter
case.  

Thus the NLL approximation for the Wilson coefficient is given by the sum
\be
\left(D_{S^2,s}^{(2)}(\nu)\right)^{NLL}=\left(D_{S^2,s}^{(2)}(\nu)\right)^{LL}+\left(\delta
  D_{S^2,s}^{(2)}(\nu)\right)^{NLL}_{s}+\left(\delta
  D_{S^2,s}^{(2)}(\nu)\right)^{NLL}_{p}+\left(D_{S^2,s}^{(2)}\right)^{~}_{\rm
  1-loop}
\,.
\label{Dsum}
\ee
where $D_{S^2,s}^{(2)}(\nu)\equiv D_{S^2,s}^{(2)}(\nu,\nu)$ and we  combine the soft and potential
running  by setting $\nu_s=\nu_p=\nu$,
which is consistent at the order of interest. 
From Eqs.~(\ref{pot}) and~(\ref{momdep})  
we  obtain the final result for the NLL spin-flip potential
\be 
\delta {\cal  H}_{\rm spin} =\left[\left(D_{S^2,s}^{(2)}(\nu)\right)^{NLL}+ 
{\gamma_s^{(1)}\over\pi}\left(\alpha_s^2c_F^2\right)^{LL}\ln\left({k\over\nu}\right)\right]
{4C_F\pi\over 3m_1m_2}\bfm{S}^2
\,.
\label{finpot}
\ee


\section{Hyperfine splitting in NLL approximation}
We are now in the position to derive the 
NLL result for the HFS. It is obtained by computing the 
corrections to the energy levels with the insertion of the
potential~(\ref{finpot})
in the quantum mechanical perturbation theory.
The result for principal quantum number $n$ reads
\bea
 E_{n,{\rm hfs}}^{NLL} &=&-{16 \over 3} { C_F^2
\alpha_s\over n} {m_r^2 \over m_1m_2}E_n^C  
\left\{
(1+2\delta \phi_n)
\left(D_{S^2,s}^{(2)}(\nu)\right)^{LL}
\right.
\nn\\&&
+\left(-\ln{\left({n\nu\over \bar\nu }\right)}
+\Psi_1(n+1)+\gamma_E+{n-1 \over 2n}
\right){\gamma_s^{(1)}\over\pi}\left(\alpha_s^2c_F^2\right)^{LL}
\nn\\&&
+\left.
 \left(\delta D_{S^2,s}^{(2)}(\nu)\right)^{NLL}_{s}
+ \left(\delta D_{S^2,s}^{(2)}(\nu)\right)^{NLL}_{p}
+\left(D_{S^2,s}^{(2)}\right)^{NLL}_{\rm 1-loop}
\right\} 
\,,
\label{HFS}
\eea
where $\bar\nu=2C_F\alpha_sm_r$,
$E_n^C= - C_F^2\alpha_s^2m_r/(2n^2)$, $\Psi_n(z) =d^n \ln \Gamma
(z)/dz^n$, $\Gamma (z)$ is the
Euler $\Gamma$-function, and $\gamma_E=0.577216\ldots$ is Euler's constant. 
In Eq.~(\ref{HFS}) the first order correction to the Coulomb wave
function at the origin due to one-loop contribution to the static
potential reads \cite{KPP}
\be
\delta \phi_n ={\alpha_s \over \pi}
\left[{3\over 8}a_1+
{\beta_0\over 4}
\left(
3\ln{\left({n\nu\over \bar\nu }\right)}
+\Psi_1(n+1)-2n\Psi_2(n)-1+\gamma_E+{2\over n}
\right)
\right]
\,,
\ee
where $a_1=31C_A/9-20T_Fn_l/9$. Furthermore, the second line 
of Eq.~(\ref{HFS}) results from the second term in square brackets
in  Eq.~(\ref{finpot}) after average over the Coulomb wave 
function. 
By expanding the
resummed expression up to ${\cal O}(\alpha_s^2)$, we get
\bea
&&
E_{n,{\rm hfs}}^{NLL} =-{16 \over 3} { C_F^2 \alpha_s^2\over n} {m_r^2 \over
m_1m_2} E_n^C
\left\{
1 + {{\als \over \pi}}\,\left[ {C_F} + \frac{7\,{C_A}\,{L^n_{\alpha_s}}}{4} 
 + 
     \frac{7\,{C_A}}{8} \,\ln\left(\frac{4\,m_r^2}{ m_1 m_2 } \right)
\right.\right.\nn\\&&
+ 
     \left( \frac{-3\,{C_F}\,{m_r}}{ {m_1} - {m_2}  } + 
        \frac{3\,{C_A}\,\left( {m_1} + {m_2} \right) }{8\,\left( {m_1} - {m_2} \right) } \right) \,
      \ln\left(\frac{{m_1}}{{m_2}} \right)
+ 
     \frac{{n_f}\,{T_F}\,
        \left( -15 - 11\,{n} + 
          12\,{{n}}^2\,\Psi_2(n) \right) }{9\,{n}} 
\nn
\\
&&
\left.
-
      \frac{{C_A}\,\left( -393 - 41\,{n} - 
          126\,{\gamma_E}\,{n} - 
          126\,{n}\,\Psi_1(n) + 
          264\,{{n}}^2\,\Psi_2(n) \right) }{72\,{n}}
     \right]  
\nn
\\
&&
+ {{{\als^2 \over \pi^2}}}{{L^n_{\alpha_s}}}\left[ L^n_{\alpha_s}
      \left( \frac{19\,{{C_A}}^2}{6} - 
        \frac{5\,{C_A}\,{n_f}\,{T_F}}{6} \right)
\right.\nn\\&&
+ 
      \left( \frac{-{{C_A}}^2}{6} - 
           \frac{11\,{C_A}\,{C_F}}{8} - 
           \frac{{{C_F}}^2\, 2 m_1 m_2  }{{\left( {m_1} + {m_2} \right) }^2} \right) \,{\pi }^2 - 
        \frac{2\,{C_F}\,{n_f}\,{T_F}}{3} 
\nn\\&&
+ 
        \left( \frac{11\,{{C_A}}^2\,\left( {m_1} + {m_2} \right) }{8\,\left( {m_1} - {m_2} \right) } + 
           \frac{4\,{C_F}\,{n_f}\,{T_F}\,{m_r}}{{m_1} - {m_2} } + 
           {C_A}\,\left( \frac{-11\,{C_F}\,{m_r}}{{m_1} - {m_2} } 
\right.\right.\nn\\&&\left.\left.
- 
              \frac{{n_f}\,{T_F}\,\left( {m_1} + {m_2} \right)}{2\,\left( {m_1} - {m_2} \right) } \right)  \right) \,
         \ln\left(\frac{{m_1}}{{m_2}} \right) 
+ 
        \left( \frac{19\,{{C_A}}^2}{6} - 
           \frac{5\,{C_A}\,{n_f}\,{T_F}}{6} \right) \,
         \ln\left(\frac{4\,{m_r^2}}{m_1 m_2}\right) 
\nn\\&&
- 
        \frac{{{C_A}}^2\,\left( -1380 - 305\,{n} - 
             450\,{\gamma_E}\,{n} - 
             450\,{n}\,\Psi_1(n) + 
             924\,{{n}}^2\,\Psi_2(n) \right) }{144\,
           {n}} 
\nn\\&&\left.\left.
+ {C_A}\,
         \left( \frac{43\,{C_F}}{12} + 
           \frac{{n_f}\,{T_F}\,
              \left( -114 - 109\,{n} - 18\,{\gamma_E}\,{n} - 
                18\,{n}\,\Psi_1(n) + 
                84\,{{n}}^2\,\Psi_2(n) \right) }{36\,
              {n}} \right)  \right]
\right\}
\label{ser}
\,,
\eea
where $\alpha_s\equiv \alpha_s(\nu)$, $\Psi_n(x)=d^n\ln\Gamma(x)/dx^n$, 
$L^n_{\alpha_s}=\ln\left(C_F\alpha_s/n\right)$ and $\nu_h=2m_r$ and
$\nu=\bar{\nu}/n$ has been chosen.  The ${\cal
O}(\alpha_s^2\ln^2\alpha_s)$ term is known \cite{Pin,HMS}, while the ${\cal
O}(\alpha_s^2\ln\alpha_s)$ term is new.
The equal-mass case expression \cite{KPPSS}, relevant for charmonium and
bottomonium, can be deduced from Eq. (\ref{HFS}) by replacing 
\be
\left(D_{S^2,s}^{(2)}\right)_{\rm 1-loop}\to
\left(D_{S^2,s}^{(2)}\right)^{q\bar q}_{\rm 1-loop}
\ee
and setting $m_1=m_2$. After including the one-photon annihilation
contribution, the  Abelian part of the equal-mass 
result 
reproduces the ${\cal O}(m\alpha_s^6\ln\alpha_s)$ and
${\cal O}(m\alpha_s^7\ln^2\alpha_s)$ corrections to the positronium 
HFS (see {\it e.g.} \cite{Pac,CMY}).

\section{Numerical estimates and conclusions}
\begin{figure}[t]
\begin{center}
\epsfxsize=\textwidth
\epsffile{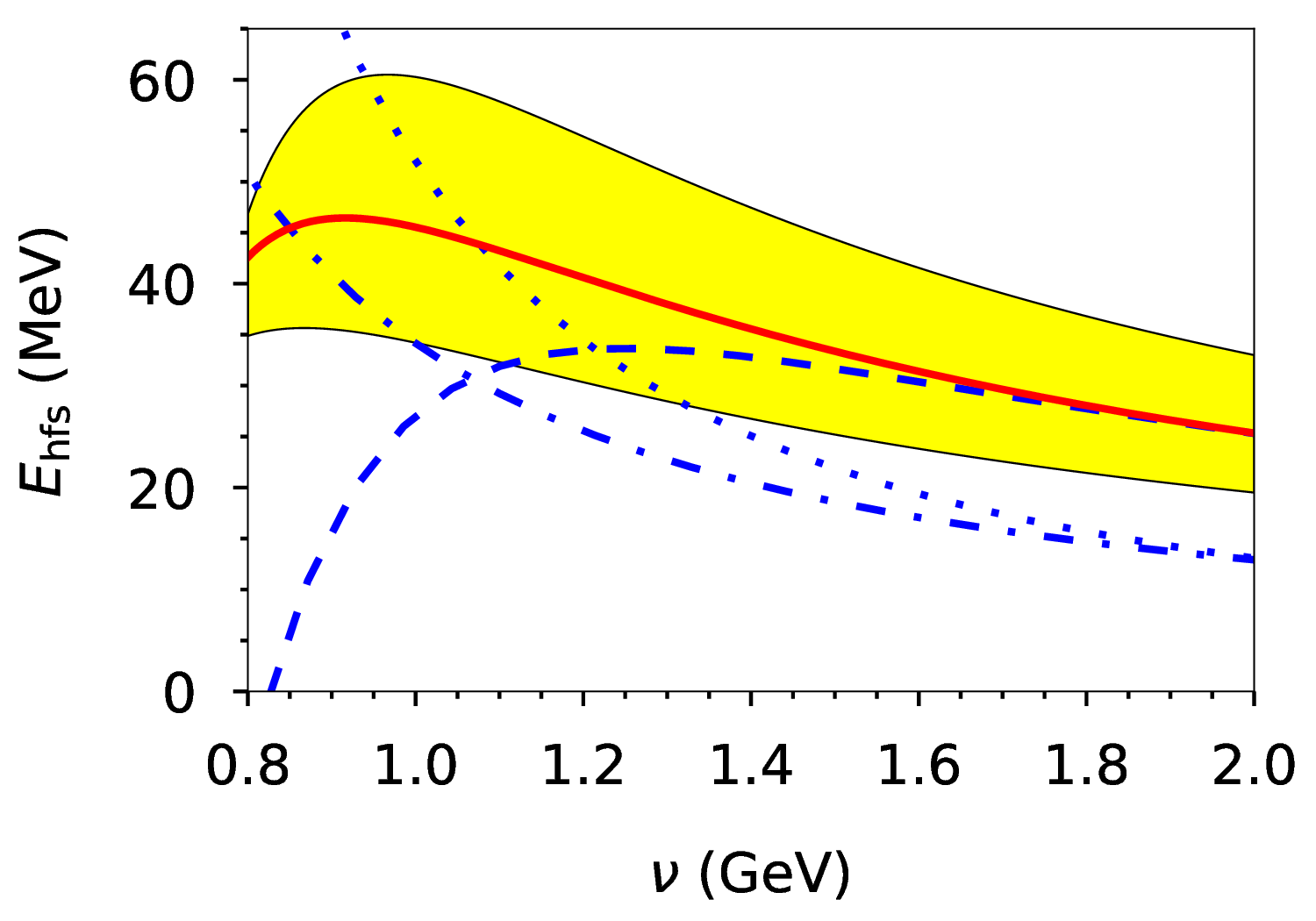}
\end{center}
\caption{\label{fig1} HFS for charm-bottom quarkonium as the function of the
renormalization scale $\nu$ in LO (dotted line), NLO (dashed line), LL
(dot-dashed line), and NLL (solid line) approximation for $\nu_h=2.05$~GeV.
For the NLL result the band reflects the errors due to $\alpha_s(M_Z)=0.118\pm
0.003$.}
\end{figure}

\begin{figure}[t]
\begin{center}
\epsfxsize=\textwidth
\epsffile{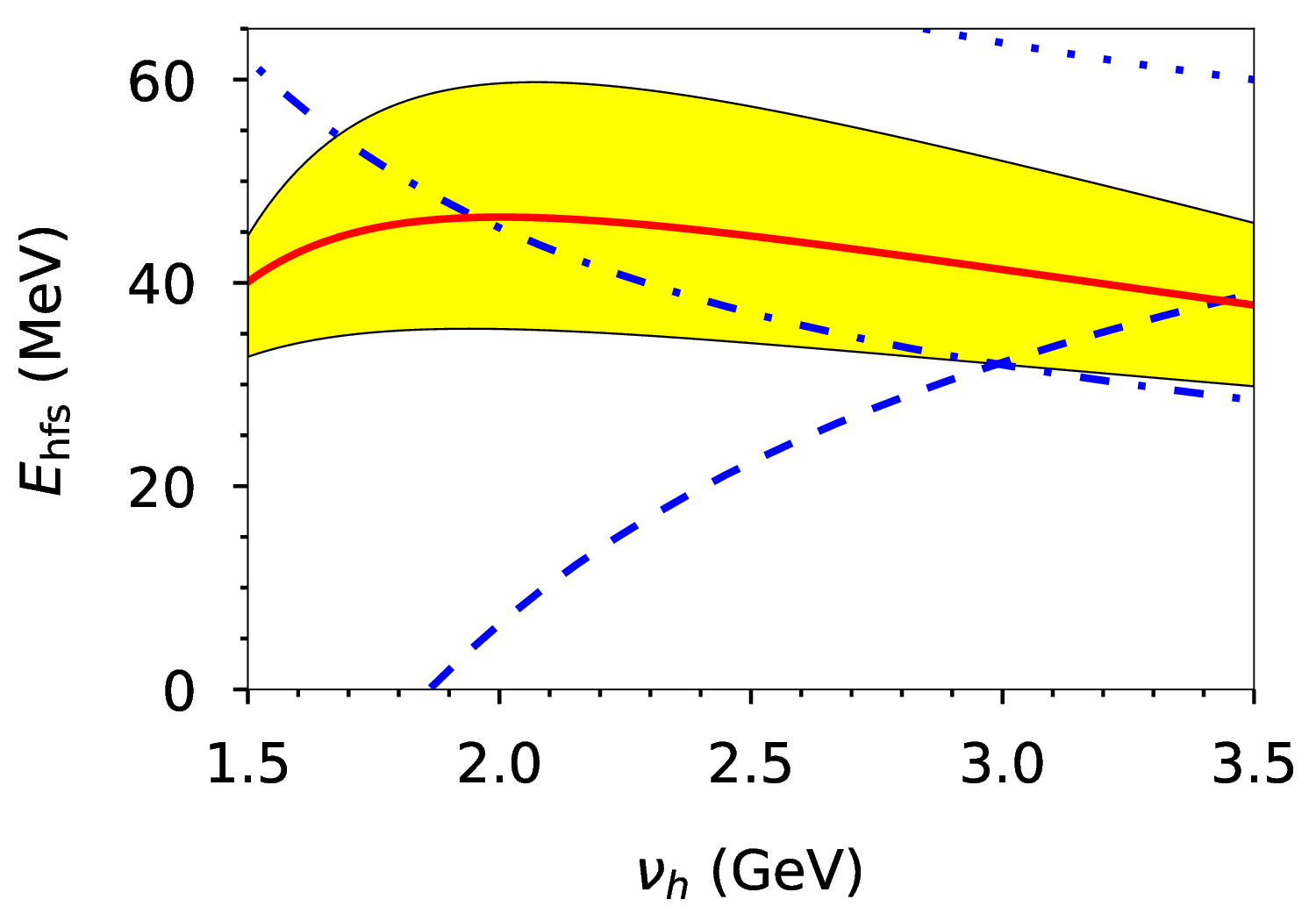}
\end{center}
\caption{\label{fig2} HFS for charm-bottom quarkonium as the function of the
hard matching scale $\nu_h$ in LO (dotted line), NLO (dashed line), LL
(dot-dashed line), and NLL (solid line) approximation for $\nu=0.915$~GeV. For
the NLL result the band reflects the errors due to $\alpha_s(M_Z)=0.118\pm
0.003$.}
\end{figure}

For the numerical estimates, we adopt the strategy of \cite{KPPSS} and replace
the on-shell mass of the charm and bottom quarks by one half of the physical
masses of the ground state of bottomonium and charmonium~\cite{Hag}. In
practice, we take $m_b=4.73$ GeV and $m_c=1.5$ GeV, consistent with the
accuracy of our computation.  Furthermore, we take $\alpha_s(M_Z)$ as an input
and run\footnote{For the running and decoupling of $\alpha_s$ we 
  use the program {\tt RunDec}~\cite{Chetyrkin:2000yt}.}
with four-loop accuracy down to the matching scale $\nu_h$ to ensure
the best precision.  Below the matching scale the running of $\alpha_s$ is
used according to the logarithmic precision of the calculation in order not to
include next-to-next-to-leading logarithms in our analysis.  In
Fig.~\ref{fig1}, the HFS for the charm-bottom quarkonium ground state is
plotted as a function of $\nu$ in the LO, NLO, LL, and NLL approximations for
the hard matching scale value $\nu_h= 2.05$~GeV.  As we see, the LL curve
shows a weaker scale dependence compared to the LO one.  The scale dependence
of the NLO and NLL expressions is further reduced, and, moreover, the NLL
approximation remains stable at the physically motivated scale of the inverse
Bohr radius, $C_F\alpha_s m_r\sim 0.9$~GeV, where the fixed-order expansion
breaks down. At the scale $\nu'\approx 0.85$~GeV, which is close to the inverse
Bohr radius, the NLL correction vanishes.  Furthermore, at $\nu''=
0.915$~GeV, the result becomes independent of $\nu$; {\it i.e.}, the NLL curve
shows a local maximum corresponding to $E_{\rm hfs}=46$~MeV, which we take as
the central value of our estimate.  The NLL curve also shows an impressive
stability with respect to the hard matching scale variation in the physical
range $m_c<\nu_h<m_b$, as we observe in Fig.~\ref{fig2}. The NLL curve has a
local maximum very near $\nu_h= 2.05$~GeV, which we take for the numerical estimates.
All this suggests a nice convergence of the logarithmic expansion despite the
presence of the ultrasoft contribution where $\alpha_s$ is normalized at the
rather low scale $\bar\nu^2/\nu_h\sim 0.5$~GeV.

Let us discuss the accuracy of our result.  For a first estimate of the error
due to uncalculated higher-order contributions, we take $7$~MeV, the
difference of the NLL and LL results at the local maxima.  A different
estimate can be obtained by varying the normalization scale in the physical
range $0.8\le\nu\le 1.4$~GeV. In this case the difference with the maximum is
$11$~MeV. Being conservative, we take this second number for our estimate of
the perturbative error. Within the power counting assumed in this paper, the
nonperturbative effects are beyond the accuracy of our computation and should
be added to the errors. Following~\cite{KPPSS}, we infer them using charmonium
data. For an estimate we attribute the whole difference between perturbation
theory and the experimental result, $\approx 5$ MeV, to nonperturbative
effects. The nonperturbative contribution in heavy quarkonium is suppressed by
the quark masses at least as $1/(m_1m_2m_r)$ and  should be smaller 
for the charm-bottom bound state.  We, however,  take $10$~MeV
as a conservative estimate  of the nonperturbative contribution to the HFS in
$B_c$. A further uncertainty is introduced by the error of $\alpha_s(M_Z)$.
In Figs.~\ref{fig1} and~\ref{fig2} this is reflected by the yellow
band, which is based on $\alpha_s(M_Z)=0.118\pm0.003$.
At the scale $\nu''= 0.915$~GeV it induces an uncertainty of
${}^{+13}_{-11}$~MeV.

To conclude, we have computed the HFS for a heavy quarkonium composed of
quark and antiquark of different flavors in the NLL approximation by
summing up the subleading logarithms $\alpha_s^n\ln^{n-1}\alpha_s$ to all
orders in the perturbative expansion.  The use of the NRG stabilizes the
result with respect to the $\nu$ variation at the 
physical scale of the inverse Bohr
radius and allow for solid first principle theoretical predictions.  An
explicit result for the two-loop soft anomalous dimension of the spin-flip
potential is also presented.

We predict the mass splitting of the vector and pseudoscalar
charm-bottom mesons 
\be
M(B^*_c)-M(B_c)=46 \pm
15\,{(\rm th)}\,{}^{+13}_{-11}\,(\delta\alpha_s)~{\rm MeV}
\ee
where the errors due to the high-order perturbative corrections and the
nonperturbative effects are added up in quadrature in ``th'', whereas
``$\delta\alpha_s$'' stands for the uncertainty in
$\alpha_s(M_Z)=0.118\pm0.003$.  With improving statistics and precision of the
$B_c$ data our result can be considered as a prediction for the $B_c^*$ meson
mass.


\vspace{5mm}
\noindent
{\bf Acknowledgments:}\\ 
We thank Bernd Kniehl for carefully reading the manuscript and useful
comments. The work of A.A.P. was supported in part by BMBF
Grant No.\ 05HT4VKA/3 and SFB Grant No. TR 9. The work of A.P. was supported
in part by MCyT and Feder (Spain), FPA2001-3598, by CIRIT (Catalonia),
2001SGR-00065 and by the EU network EURIDICE, HPRN-CT2002-00311. The work of
V.A.S. was supported in part by RFBR Project No. 03-02-17177, Volkswagen
Foundation Contract No. I/77788, and DFG Mercator Visiting Professorship
No. Ha 202/1. M.S. was supported by HGF Grant No. VH-NH-008.



\section*{Appendix}

The analytical results for the coeffcients $f_i$ and $A_i$ of 
Eq.~(\ref{DNLLp}) read
($z=\left(\alpha_s(\nu_p)/\alpha_s(\nu_h)\right)^{1/\beta_0}$)  
\begin{eqnarray}
  && 
  f_{1} = z^{3 \betnull - 2 C_A}   {}_2F_1\left(3 - \frac{2 C_A}{\betnull}, 1; 4
- \frac{2 C_A}{\betnull}; \frac{z^{\betnull}}{2}\right)\,,\quad
  f_{2} = z^{2 \betnull - (25 C_A)/6}\,,\quad
  \nonumber\\
  &&
  f_{3} = z^{2 \betnull - 4 C_A}\,,\quad
  f_{4} = z^{2 \betnull - 3 C_A}\,,\quad
  f_{5} = z^{2 \betnull - 2 C_A}\,,\quad
  f_{6} = z^{2 \betnull - 2 C_A}\logminus \,,\quad
  \nonumber\\
  &&
  f_{7} = z^{2 \betnull - C_A}\,,\quad
  f_{8} = z^{\betnull - (13 C_A)/6}\,,\quad
  f_{9} = z^{\betnull - 2 C_A}\,,\quad
  f_{10} = z^{2 \betnull}\,,\quad
  \nonumber\\
  &&
  f_{11} =
B_{z^{\beta_0}/2}\left(2-\frac{2C_A}{\betnull},1+\frac{2C_A}{\betnull}\right),
\quad
  f_{12} = z^{\betnull}\,,\quad
  f_{13} = z^{\betnull}\logminus\,,\quad
  \nonumber\\
  &&
  f_{14} = \ln(z)\,,\quad
  f_{15} = 1\,,\quad
  f_{16} = \logminus \,,
\end{eqnarray}
\begin{eqnarray}
  A_{1} &=&
  \frac{C_F \left(C_A^2+3 C_F C_A+2 C_F^2\right) \left(C_A-8
   n_l T_F\right)}{2 \left(5 C_A-4 n_l T_F\right) \left(9 C_A-4
   n_l T_F\right) \left(2 C_A-n_l T_F\right)}
  \,,\nonumber\\
  A_{2} &=&
  \frac{192 C_F^2
   \left(5 C_A+8 C_F\right) n_l T_F \left(C_A-8 n_l T_F\right) (2\mu
   _r-1)}{13 C_A \left(19 C_A-16 n_l T_F\right) \left(9 C_A-8 n_l
   T_F\right) \left(5 C_A-4 n_l T_F\right)}
  \,,\nonumber\\
  A_{3} &=&
  \frac{3 C_F^2 \left(63 C_A^2-264
   n_l T_F C_A+240 n_l^2 T_F^2\right) \mu _r}{4 \left(5 C_A-4 n_l
   T_F\right){}^3}-\frac{3 C_F^2 \left(C_A-8 n_l T_F\right)}{2
   \left(5 C_A-4 n_l T_F\right){}^2}
  \,,\nonumber\\
  A_{4} &=&
  \frac{
    -3 C_A C_F
  }{
    4 (13 C_A - 8 n_l T_F)}
  \,,\nonumber\\
  A_{5} &=&
  \frac{3}{208
   C_A \left(5 C_A-4 T_F n_l\right) \left(11
   C_A-4 T_F n_l\right) \left(2 C_A-T_F
   n_l\right)} 
\nn
\\
&&
\times
\left(26 C_A^3 \left(-284 C_F T_F
   n_l+C_F^2 \left(4 \mu _r-269\right)-8 T_F^2
   n_l^2\right)
\right.
\nn
\\
&&
+16 C_A^2 C_F \left(C_F T_F n_l
   \left(439-59 \mu _r\right)+13 C_F^2 \left(7
   \mu _r-2\right)+130 T_F^2 n_l^2\right)
\nn
\\
&&
-32
   C_A C_F^2 T_F n_l \left(4 C_F \left(95 \mu
   _r-28\right)+T_F n_l \left(53-28 \mu
   _r\right)\right)+416 C_A^4 \left(11 C_F+2
   T_F n_l\right)
\nn
\\
&&
\left.
-715 C_A^5+2048 C_F^3 T_F^2
   n_l^2 \left(2 \mu _r-1\right)\right)
\,,
  \nonumber\\
  A_{6} &=&
    \frac{3 C_F \left(C_A^2+3 C_F C_A+2
   C_F^2\right) \left(C_A-8 n_l T_F\right)}{\left(5 C_A-4 n_l
   T_F\right) \left(11 C_A-4 n_l T_F\right) \left(2 C_A-n_l
   T_F\right)}
  \,,\nonumber\\
  A_{7} &=&
  \frac{
    -3 (C_A - 3 C_F) C_F
  }{
    19 C_A - 8 n_l T_F}
  \,,\nonumber\\
  A_{8} &=&
  -\frac{1728 C_F^2 \left(5 C_A+8 C_F\right) n_l
   T_F(1-2\mu_r)}{13 \left(9 C_A-8 n_l T_F\right){}^2 \left(5 C_A-4 n_l
   T_F\right)}
  \,,\nonumber\\
  A_{9} &=&
  \frac{9 C_F^2 }{\left(9 C_A-8 T_F
   n_l\right) \left(5 C_A-4 T_F n_l\right){}^3
   \left(11 C_A-4 T_F n_l\right)}
\nn
\\
&&
\times
\left(8 C_A^3 \left(15 C_F
   \left(2-7 \mu _r\right)+T_F n_l
   \left(107-781 \mu _r\right)\right)+16 C_A^2
   T_F n_l \left(2 C_F \left(241 \mu
   _r-66\right)
\right.\right.
\nn
\\
&&
\left.\left.+T_F n_l \left(302 \mu
   _r+17\right)\right)-128 C_A T_F^2 n_l^2
   \left(4 C_F \left(16 \mu _r-3\right)+T_F n_l
   \left(10 \mu _r+1\right)\right)
\right.
\nn
\\
&&
\left.
+3 C_A^4
   \left(992 \mu _r-415\right)+2048 C_F T_F^3
   n_l^3 \mu _r\right)
  \,,\nonumber\\
  A_{10} &=&
  \mrM \frac{
    3 C_F^2 
  }{
    4 (11 C_A - 4 n_l T_F)}
  \,,\nonumber\\
   A_{11} &=&
  \frac{
    6C_A(C_A-2C_F)
  }{
     (11 C_A - 4 n_l T_F)}
  \,,\nonumber\\
  A_{12} &=&
  \frac{27 C_F }{13 \left(5 C_A-4 T_F
   n_l\right) \left(11 C_A-4 T_F
   n_l\right){}^2}
\nn
\\
&&
\times
\left(26 C_A^2 \left(C_F \left(2
   \mu _r-9\right)-2 T_F n_l\right)+8 C_A C_F
   \left(13 C_F \left(7 \mu _r-4\right)-T_F n_l
   \left(7 \mu _r+3\right)\right)
\right.
\nn
\\
&&
\left.+39 C_A^3+128
   C_F^2 T_F n_l \left(1-2 \mu
   _r\right)\right)
  \,,\nonumber\\
  A_{13}&=&
 \frac{216 C_A C_F \left(C_A+C_F\right)
   \left(C_A+2 C_F\right)}{\left(5 C_A-4 T_F
   n_l\right) \left(11 C_A-4 T_F
   n_l\right){}^2}
  \,,\nonumber\\
  A_{14} &=&
  \frac{216 C_A C_F^2 }{\left(9 C_A-8 T_F
   n_l\right) \left(5 C_A-4 T_F n_l\right){}^2
   \left(11 C_A-4 T_F n_l\right)}
\nn
\\
&&
\times
\left(\mu _r \left(2 C_A
   T_F n_l \left(9 C_A+62 C_F\right)-3 C_A^2
   \left(8 C_A+35 C_F\right)-32 C_F T_F^2
   n_l^2\right)
\right.
\nn
\\
&&
\left.
+2 C_A \left(5 C_A-4 T_F
   n_l\right) \left(3 \left(C_A+C_F\right)-T_F
   n_l\right)\right)
  \,,\nonumber\\
  A_{15} &=& -  {}_2F_1\left(1, 1; 4 - \frac{2 C_A}{\betnull}; -1\right)\frac{
    C_F (C_A + 2 C_F) (C_A - 8 n_l T_F)
  }{
    (5 C_A - 4 n_l T_F) (9 C_A - 4 n_l T_F) (2 C_A - n_l T_F)}
  \left(C_A + 4C_F \mrM\right)
  \nonumber\\&&\mbox{}
 -B_{1/2}\left(2-\frac{2C_A}{\betnull},1+\frac{2C_A}{\betnull}\right) \frac{
    6C_A(C_A-2C_F)
  }{
     (11 C_A - 4 n_l T_F)}+
\frac{3C_A(C_A-2C_F) }{16 (2C_A -  n_l T_F) 
    }
  \nonumber\\&&\mbox{}
  +\frac{
    - 3 C_A C_F}{8 (13 C_A - 8 n_l T_F) 
    (19 C_A - 8 n_l T_F) (5 C_A - 4 n_l T_F) (11 C_A - 4 n_l T_F)^2 (2 C_A -
  n_l T_F)} \nonumber\\&& \mbox{}\qquad
  \times \left(
263641 C_A^5 - 919114 C_A^4 n_l T_F 
    + 1071256 C_A^3 n_l^2 T_F^2 - 556448 C_A^2 n_l^3 T_F^3 
\right.
\nonumber\\&& \mbox{}\qquad\qquad
\left.
+ 131456 C_A n_l^4 T_F^4 -
    11264 n_l^5 T_F^5\right.)   
  \nonumber\\&&\mbox{}
  +\frac{
    27 C_A C_F^3 }{(9 C_A - 8 n_l T_F)^2 
    (19 C_A - 16 n_l T_F) (5 C_A - 4 n_l T_F)^2 (11 C_A - 4 n_l T_F)^2}
  \nonumber\\&&\mbox{}
  \times \frac{1}{ (7 C_A - 2 n_l T_F) (2 C_A - n_l T_F)}
    \left(3644181 C_A^6 - 7690472 C_A^5 n_l T_F + 3453968 C_A^4 n_l^2 T_F^2 
  \right.
  \nonumber\\&&\mbox{}\qquad\qquad 
  \left. +
    3026560 C_A^3 n_l^3 T_F^3 - 3419648 C_A^2 n_l^4 T_F^4 
    + 1150976 C_A n_l^5 T_F^5 - 131072 n_l^6 T_F^6 \right) 
  \nonumber\\&&\mbox{}
  +\frac{
     3 C_F^2}{16 (19 C_A - 16 n_l T_F) (9 C_A - 8 n_l T_F)^2 (19 C_A - 
    8 n_l T_F) (5 C_A - 4 n_l T_F)^2}
  \nonumber\\&&\mbox{}\quad
  \times\frac{1}{ (11 C_A - 4 n_l T_F)^2  (7 C_A - 
    2 n_l T_F) (2 C_A - n_l T_F)}
 \left(12488524839 C_A^9 \right.
  \nonumber\\&&\mbox{}\quad\quad
  - 37966954860 C_A^8 n_l T_F 
    + 37940834480 C_A^7 n_l^2 T_F^2 -
    1336115840 C_A^6 n_l^3 T_F^3 
  \nonumber\\&&\mbox{}\quad\quad
- 27950404608 C_A^5 n_l^4 T_F^4 
    + 25870953472 C_A^4 n_l^5 T_F^5 
    - 11448205312 C_A^3 n_l^6 T_F^6 
  \nonumber\\&&\mbox{}\quad\quad
  \left.
    + 2764505088 C_A^2 n_l^7 T_F^7 
    - 343932928 C_A n_l^8 T_F^8 + 16777216 n_l^9 T_F^9\right)
  \nonumber\\&&\mbox{}
  +\mrM\left[ \frac{
      -3 C_F^3       }{(19 C_A - 16 n_l T_F) (9 C_A - 8 n_l T_F)^2 (5 C_A - 4
  n_l T_F)^2 (11 C_A - 4 n_l T_F)^2}
    \right.
  \nonumber\\&&\mbox{}\quad
  \times\frac{1}{ (7 C_A - 2 n_l T_F) (2 C_A - n_l T_F)}
  (62685009 C_A^7 - 91230606 C_A^6 n_l T_F 
  \nonumber\\&&\mbox{}\quad\quad
  -78455168 C_A^5 n_l^2 T_F^2 + 233772512 C_A^4 n_l^3 T_F^3 
      - 176816384 C_A^3 n_l^4 T_F^4 
  \nonumber\\&&\mbox{}\quad\quad
  + 58415104 C_A^2 n_l^5 T_F^5 
      - 7979008 C_A n_l^6 T_F^6 + 262144 n_l^7 T_F^7) 
  \nonumber\\&& \mbox{}
    +\frac{
      - 3 C_F^2 }{4 (19 C_A - 16 n_l T_F) (9 C_A - 8 n_l T_F)^2 (5 C_A - 4 n_l
  T_F)^3 (11 C_A - 4 n_l T_F)^2} 
  \nonumber\\&&\mbox{}\quad
  \times \frac{1}{ (7 C_A - 2 n_l T_F) (2 C_A - n_l T_F)}(659490741 C_A^9 -
  1386410130 C_A^8 n_l T_F 
  \nonumber\\&&\mbox{}\quad\quad
      - 876382076 C_A^7 n_l^2 T_F^2 + 5528200720 C_A^6 n_l^3 T_F^3 
      - 7422517824 C_A^5 n_l^4 T_F^4 
  \nonumber\\&&\mbox{}\quad\quad
  + 5156251904 C_A^4 n_l^5 T_F^5 - 2102788096 C_A^3 n_l^6 T_F^6 +
      511131648 C_A^2 n_l^7 T_F^7 
  \nonumber\\&&\mbox{}\quad\quad
  - 69730304 C_A n_l^8 T_F^8 + 4194304 n_l^9
  T_F^9)  
  \Bigg]
\nn
\\
&&
+
\frac{C_F^2 \left(4 \mu _r-1\right)}{16 \left(19 C_A-16 T_F
   n_l\right) \left(9 C_A-8 T_F n_l\right)
   \left(5 C_A-4 T_F n_l\right) \left(9 C_A-4
   T_F n_l\right) \left(11 C_A-4 T_F
   n_l\right){}^2}
\nn
\\
&&
\times 
\frac{1}{
\left(7 C_A-2 T_F n_l\right)
   \left(2 C_A-T_F n_l\right)}
\left(8
   \left(C_A+2 C_F\right) \left(19 C_A-16 T_F
   n_l\right) \left(C_A-8 T_F n_l\right)
\right.
\nn
\\
&&
\left.
   \left(9 C_A-8 T_F n_l\right) \left(11 C_A-4
   T_F n_l\right){}^2 \left(7 C_A-2 T_F
   n_l\right) \, _2F_1\left(1,-\frac{2 C_A}{\beta_0}+3;-\frac{2 C_A}{\beta_0}+4;\frac{1}{2}\right)
\right.
\nn
\\
&&
\left.
+331018056 C_A^6
   T_F n_l-598155936 C_A^5 T_F^2 n_l^2-96081984
   C_A^5 C_F T_F n_l+541880448 C_A^4 T_F^3
   n_l^3
\right.
\nn
\\
&&
\left.
+112001280 C_A^4 C_F T_F^2
   n_l^2-273654528 C_A^3 T_F^4 n_l^4-93545472
   C_A^3 C_F T_F^3 n_l^3+78680064 C_A^2 T_F^5
   n_l^5
\right.
\nn
\\
&&
\left.
+50577408 C_A^2 C_F T_F^4
   n_l^4-12140544 C_A T_F^6 n_l^6-14352384 C_A
   C_F T_F^5 n_l^5+42033168 C_A^6 C_F
\right.
\nn
\\
&&
\left.
-66997287
   C_A^7+1572864 C_F T_F^6 n_l^6+786432 T_F^7
   n_l^7\right)
\,,
  \nonumber\\
  A_{16}&=&
   -\frac{432 C_A C_F \left(C_A+C_F\right)
   \left(C_A+2 C_F\right)}{\left(5 C_A-4 T_F
   n_l\right) \left(11 C_A-4 T_F
   n_l\right){}^2}
  \,, \nonumber\\&&
\end{eqnarray}
with $\mrM=m_r/(m_1+m_2)$, $B_{z}(a,b)$ is the
incomplete beta-function,
and ${}_2F_1(a,b;c;z)$ is the
hypergeometric function.


\begin{thebibliography}{99}

\bibitem{Abe} CDF Collaboration, F. Abe  {\it et al.},
Phys.\ Rev.\ Lett.\  {\bf 81}, 2432 (1998).


\bibitem{Eichten:1994gt}
E.~J.~Eichten and C.~Quigg,
Phys.\ Rev.\ D {\bf 49}, 5845 (1994).

\bibitem{Gershtein:1994jw}
S.~S.~Gershtein, V.~V.~Kiselev, A.~K.~Likhoded and A.~V.~Tkabladze,
Phys.\ Usp.\  {\bf 38}, 1 (1995)
[Usp.\ Fiz.\ Nauk {\bf 165}, 3 (1995)].

\bibitem{Sha} UKQCD Collaboration, H.P. Shanahan {\it et al.},
Phys.\ Lett.\ B {\bf 453}, 289 (1999).

\bibitem{BraVai} N. Brambilla and A. Vairo,
Phys.\ Rev.\ D {\bf62}, 094019 (2000)

\bibitem{BucNg} 
W. Buchm\"uller, Y.J. Ng, and S.H.H. Tye,  
Phys.\ Rev.\ D {\bf24}, 3003 (1981);
S.N. Gupta, S.F. Radford, and W.W. Repko 
Phys.\ Rev.\ D {\bf26}, 3305 (1982);
J. Pantaleone, S.H.H. Tye, and Y.J. Ng,  
Phys.\ Rev.\ D {\bf33}, 777 (1986).

\bibitem{recent}
S. Titard and F.J. Yndurain,
Phys.\ Rev.\ D {\bf49}, 6007 (1994),
A. Pineda and F.J. Yndurain,
Phys.\ Rev.\ D {\bf 58},  094022 (1998);
A.A. Penin and A.A. Pivovarov,
Nucl.\ Phys.\ {\bf B550}, 375 (1999);
Yad.\ Fiz.\ {\bf64}, 323 (2001)
[Phys.\ Atom.\ Nucl.\ {\bf64}, 275 (2001)];
S. Recksiegel and Y. Sumino,
Phys.\ Lett.\ B {\bf578}, 369 (2004).

\bibitem{PenSte} A.A. Penin and M. Steinhauser,
Phys.\ Lett.\ B {\bf538}, 335 (2002).

\bibitem{KPPSS} B.A. Kniehl, A.A. Penin, 
A. Pineda, V.A. Smirnov, and M. Steinhauser, 
Report No. DESY-03-172, TTP-03-40, UB-ECM-PF-03-28, and hep-ph/0312086.

\bibitem{CasLep} W.E. Caswell and G.P. Lepage,
Phys.\ Lett.\ B {\bf167}, 437 (1986);
G.T. Bodwin, E. Braaten, and G.P. Lepage,
Phys.\ Rev.\ D {\bf51}, 1125 (1995); {\bf55}, 5853(E) (1997).

\bibitem{PinSot1} A. Pineda and J. Soto,
Nucl.\ Phys.\ B (Proc.\ Suppl.) {\bf64}, 428 (1998);
B.A. Kniehl and A.A. Penin,
Nucl.\ Phys.\ {\bf B563}, 200 (1999);
N. Brambilla, A. Pineda, J. Soto, and A. Vairo,
Nucl.\ Phys.\ {\bf B566}, 275 (2000).

\bibitem{Pin} A. Pineda,
Phys.\ Rev.\ D {\bf65}, 074007 (2002); 
{\bf66}, 054022 (2002).

\bibitem{HMS} A.~V.~Manohar and I.~W.~Stewart,
Phys.\ Rev.\ D {\bf 62}, 014033 (2000); 
A.H. Hoang and I.W. Stewart,
Phys.\ Rev.\ D {\bf 67}, 114020 (2003).

\bibitem{BenSmi} M. Beneke and V.A. Smirnov,
Nucl.\ Phys.\ {\bf B522}, 321 (1998);
V.A. Smirnov,
{\it Applied Asymptotic Expansions in Momenta and Masses}
(Springer-Verlag, Heidelberg, 2001).

\bibitem{LMR} M.E. Luke, A.V. Manohar, and I.Z. Rothstein,
Phys.\ Rev.\ D {\bf61}, 074025 (2000).

\bibitem{ABN} G. Amor\'os, M. Beneke, and M. Neubert, 
Phys.\ Lett.\ B {\bf 401}, 81 (1997).

\bibitem{PinSot2} A. Pineda and J. Soto,
Phys.\ Lett.\ B {\bf420}, 391 (1998);
Phys.\ Rev.\ D {\bf59}, 016005 (1999).

\bibitem{CMY} A. Czarnecki, K. Melnikov, and A. Yelkhovsky,
Phys.\ Rev.\ A {\bf59}, 4316 (1999).

\bibitem{KPSS} B.A. Kniehl, A.A. Penin, 
V.A. Smirnov, and M. Steinhauser, 
Phys.\ Rev.\ D {\bf65}, 091503(R) (2002);
Nucl.\ Phys.\ {\bf B635}, 357 (2002);
Phys.\ Rev.\ Lett.\ {\bf90}, 212001 (2003); 
{\bf91}, 139903(E) (2003).

\bibitem{Bai}
P.A. Baikov, Phys.\ Lett.\ B {\bf385}, 404 (1996);
Nucl.\ Instrum.\ Meth.\ A {\bf 389}, 347 (1997);
V.A. Smirnov and M. Steinhauser, 
Nucl.\ Phys.\ {\bf B672} 199 (2003).

\bibitem{KniPen2} B.A. Kniehl and A.A. Penin,
Nucl.\ Phys.\ {\bf B577}, 197 (2000).

\bibitem{Hil}  R.J. Hill,
Phys.\ Rev.\ Lett.\ {\bf86}, 3280 (2001);
K. Melnikov and A. Yelkhovsky,
Phys.\ Rev.\ Lett.\ {\bf86}, 1498 (2001);
B.A. Kniehl and A.A. Penin,
Phys.\ Rev.\ Lett.\ {\bf85}, 1210 (2000); {\bf85}, 3065(E) (2000);
{\bf85}, 5094 (2000).

\bibitem{Pac} K. Pachucki, Phys.\ Rev.\ A {\bf56}, 297 (1997).

\bibitem{Manohar} A.V. Manohar,  
Phys.\ Rev.\ D {\bf 56}, 230 (1997).
 
\bibitem{KPP} J.H. K\"uhn, A.A. Penin, and A.A. Pivovarov,
Nucl.\ Phys.\ {\bf B534}, 356 (1998);
A.A. Penin and A.A. Pivovarov,
Phys.\ Lett.\ B {\bf435}, 413 (1998);
Nucl.\ Phys.\ {\bf B549}, 217 (1999);
K. Melnikov and A. Yelkhovsky,
Phys.\ Rev.\ D {\bf59}, 114009 (1999).


\bibitem{Hag} K. Hagiwara {\it et al.},
Phys.\ Rev.\ D {\bf 66}, 010001 (2002).

\bibitem{Chetyrkin:2000yt}
K.~G.~Chetyrkin, J.~H.~Kuhn and M.~Steinhauser,
Comput.\ Phys.\ Commun.\  {\bf 133} (2000) 43.

\end{thebibliography}
\end{document}